\documentstyle[preprint,tighten,aps]{revtex}

\begin{document} 
\draft 
\preprint{
\begin{tabular}{r} FTUV/97$-$49 \\ IFIC/97$-$65
\end{tabular}
}
\title{Leptophobic character of the $Z'$ \\ in an $SU(3)_C \otimes 
SU(3)_L \otimes U(1)_X$ model} 
\author{D.\ G\'omez Dumm\thanks{e-mail: dumm@gluon.ific.uv.es}} 
\address{Departament de F\'{\i}sica Te\`orica, Universitat
de Val\`encia, \\ 
%and IFIC, Centre Mixt Univ.\ Valencia-CSIC,\\ 
c.\ Dr.\ Moliner 50, E-46100 Burjassot (Val\`encia), Spain} 
 
\maketitle 
 
\thispagestyle{empty} 
 
\begin{abstract} 
We show that the extra $Z$ boson predicted within the so-called 
``3-3-1'' model has a leptophobic character, and analyse its effects 
on $Z$ decay widths and on the $t\bar t$ production cross  
section in $p\bar p$ collisions at 
the Fermilab Tevatron. Recent model-independent analysis are applied 
in order to estimate the contribution of this $Z'$ 
to the observables that will be measured at LEP2. 
\end{abstract} 
 
 
\setcounter{page}{2} 
 
The presence of an extra neutral vector boson having relatively small 
couplings to the lepton sector (``leptophobic'') has recently been 
considered with special interest \cite{ver1,alt}. This has mainly been 
motivated by the experimental observation of possible deviations from the 
Standard Model (SM) predictions for the decay of the $Z$ boson to heavy 
quarks at LEP \cite{lep}, and the excess of jets at large $E_T$ at CDF 
\cite{cdf}. In addition, it has been noticed that such a $Z'$ boson could 
give rise to measurable effects at LEP2 and NLC, provided that the lepton 
couplings are nonvanishing \cite{ver}. 
 
The experimental situation for the mentioned LEP and CDF observations 
is still under accurate final analysis \cite{pol}. However, even in the 
absence of discrepancies 
with the SM, models including a leptophobic $Z'$ still represent an 
interesting subject, specially in connection with the on-going LEP2 
experiments. In this regard, a detailed study is performed in 
Ref.~\cite{ver}, where the possible effects of a general extra $Z$ on the 
final hadronic channels at LEP2 are calculated, and the results are 
compared with the expected experimental accuracies. It is also stressed 
that there is a phenomenologically significant difference between the 
models where the couplings among the $Z'$ and the leptons are just 
suppressed, and those where these interactions {\em exactly} vanish at the 
tree level. In the first of these two cases the presence of the $Z'$ could 
be detected at LEP2 even for $M_{Z'}\sim 1$ TeV, while a ``totally 
leptophobic'' $Z'$ is expected to yield no visible effects on the 
conventional observables. 
 
Several authors have followed the idea of a leptophobic $Z'$, and 
different models can be found in the literature, containing either 
approximate or exact suppression mechanisms for the $Z'$-lepton couplings 
\cite{otros}. The purpose of this paper is to study in this context the 
so-called ``3-3-1'' model, which was presented a few years ago 
\cite{vic,fra} following a quite different motivation. The model is 
based on an $SU(3)_C\otimes SU(3)_L\otimes U(1)_X$ gauge group, 
and attempts to explain (at least partially) the problem of family 
replication, since it has the particularity of being 
anomaly-free only if the number of lepton families 
is a multiple of the number of colors. We show here that the model 
contains a $Z'$ which falls into the leptophobic class, and analyse 
the contributions of $Z-Z'$ mixing to the ratios $R_b$, $R_c$ measured 
at LEP and the effect of $Z'$ exchange on the top quark production 
cross section in high-energy $p\bar p$ collisions. The work is 
concluded by studying the possibility of finding observable effects 
coming from the exchange of this $Z'$ in $e^+ e^-$ collisions at LEP2. 
 
Let us stress that the 3-3-1 model shows a distinctive feature, 
which is the fact that the suppression of the lepton 
couplings is not obtained from an ad hoc 
imposition, but appears as a direct consequence of the required 
fermion quantum numbers. From this point of view, the leptophobic 
character of the $Z'$ in this model can be regarded as ``natural''. 
 
\hfill 
 
{\em Outline of the model.} The structure of the 3-3-1 model has been 
detailed in several articles \cite{vic,fra,ng,cp,foot,liu}. Let us include 
here just a brief description in order to clarify the notation (we follow 
that of Ref.~\cite{vic}). 
 
The model contains the ordinary SM quarks and leptons, as well as three 
new quarks $J_1$ and $J_{2,3}$ with charges $5/3$ and $-4/3$ 
respectively. These fermions are organized into $SU(3)_L$ 
triplets and singlets as follows: 
\begin{displaymath} 
\begin{array}{c} 
\Psi_{lL} = \left( \begin{array}{l} \nu_{l} \\ l \\ l^{c} \end{array} 
\right)_{L} \sim\; (\underline{3}, 0) 
\hspace{.9cm} 
Q_{1L} = \left( \begin{array}{l} u_{1} \\ d_{1} \\ J_{1} \end{array} 
 \right)_{L} \sim\; (\underline{3}, \frac{2}{3}) \hspace{.9cm} 
Q_{2,3L} = \left( \begin{array}{l} J_{2,3} \\ u_{2,3} \\ d_{2,3} 
\end{array} \right)_{L} \sim\; (\underline{3}^{\ast}, -\frac{1}{3}) 
\nonumber \\  
\rule{0cm}{1cm} 
u_{iR} \;\sim\; (\underline{1},\frac{2}{3}) \hspace{1cm} 
d_{iR} \;\sim\; (\underline{1},-\frac{1}{3}) \hspace{1cm} 
J_{1R} \;\sim\; (\underline{1},\frac{5}{3}) \hspace{1cm} 
J_{2,3R} \;\sim\; (\underline{1},-\frac{4}{3})  
\end{array} 
\end{displaymath} 
where $l=e,\mu,\tau$, and in each case the first and second entries 
between the parentheses stand for the $SU(3)_L$ representation and $X$ 
quantum number respectively. Notice 
that, forced by the requirement of anomaly cancellation, 
one of the three quark families transforms under $SU(3)_L$ as a 
$\underline{3}$, while the other two are in a $\underline{3}^\ast$. This 
``family discrimination'' gives rise to flavor changing neutral 
interactions (FCNI) at the tree level \cite{liu,tum}. 
 
As in the SM case, the scalar sector of the model is responsible 
for the spontaneous breakdown of the gauge symmetry and the 
origin of the fermion and gauge boson masses. The scalar fields 
are organized into one $SU(3)_L$ sextuplet with $X=0$ (necessary to 
provide the lepton masses) and three $SU(3)_{L}$ triplets with 
$X$ values 1, 0 and -1 respectively. The spontaneous symmetry 
breakdown follows the hierarchy  
\begin{equation} 
SU(3)_L\otimes U(1)_X \longrightarrow\hspace{-.5cm}^V\hspace{.3cm} 
SU(2)_L\otimes U(1)_Y \longrightarrow\hspace{-.5cm}^v\hspace{.4cm} 
U(1)_{em} 
\end{equation} 
where $V$ and $v$ denote the corresponding breaking 
scales. Finally, the model includes nine gauge bosons, associated 
with the generators of the gauge group. The charged sector 
contains two relatively light singly charged particles, which can be 
identified with the usual SM $W^\pm$ bosons, and four heavier dileptons 
$Y^\pm$ and $Y^{\pm\pm}$. The neutral sector, which will be considered 
below in more detail, includes a new gauge boson $Z'$ besides the 
ordinary $Z$ and the massless photon. 
 
\hfill 
 
{\em Neutral gauge bosons and leptophobia}. As usual, the 
couplings between the fermions and the neutral gauge bosons 
can be read from the covariant derivative in the fermion kinetic 
term of the Lagrangian. One has 
\begin{eqnarray} 
{\cal L} & = &  
- \sum_{i=1}^3 \bar Q_{iL} (g' X B_\mu + g T_3 W_\mu^3 + g T_8 
W_\mu^8) \gamma^\mu Q_{iL} \nonumber \\ 
& & -\, g' \sum_{q=u_i,d_i,J_i} \bar q_R\, X B_\mu \gamma^\mu q_R  
- g \sum_{l=e,\mu,\tau} \bar\Psi_{lL} (T_3 W_\mu^3 + T_8 
W_\mu^8)\gamma^\mu \Psi_{lL}  
\label{lag} 
\end{eqnarray} 
where $B$ and $W^{3,8}$ are the gauge fields corresponding to 
the $U(1)_X$ and the diagonal $SU(3)_L$ generators respectively. 
The change to the $\{A,Z,Z'\}$ basis is obtained by means of the 
rotation 
\begin{eqnarray} 
A & = & \sin\theta\; W^3 + \cos\theta\; (-\sin\phi\; W^8 + \cos\phi\; B) 
\nonumber \\ 
Z & = & \cos\theta\; W^3 - \sin\theta\; (-\sin\phi\; W^8 + \cos\phi\; B) 
\nonumber \\ 
Z' & = & \cos\phi\; W^8 + \sin\phi\; B 
\label{rota} 
\end{eqnarray} 
where the angles $\phi$ and $\theta$ are given by 
\begin{equation} 
\cos\phi = \frac{\sqrt{1-4\sin^2\theta}}{\cos\theta} 
\;, \hspace{1.5cm} 
\frac{\sin\theta}{\sqrt{1-4\sin^2\theta}} = \frac{g'}{g} 
\label{const} 
\end{equation} 
The states $A$ and $Z$ in (\ref{rota}) can now be identified with the 
ordinary SM neutral gauge bosons. However, the $Z$ is not yet an exact 
mass eigenstate, but turns out to be slightly mixed with the $Z'$. The 
corresponding mixing angle $\beta$ is a function of the nonzero vacuum 
expectation values of the scalar fields, and it is shown to be 
suppressed by the ratio $v^2/V^2$ between the squares of the 
above mentioned symmetry breaking scales. 
 
{}From the resulting couplings between the fermions and the $Z$, it is 
seen that the angle $\theta$ corresponds to the SM Weinberg angle 
$\theta_W$. However, in contrast with the SM, the rotation (\ref{rota}) in 
the 3-3-1 model sets an upper bound for $\sin^2\theta$, namely 1/4, as it 
is seen from Eq.~(\ref{const}) \cite{fra}. The fact that the value of 
$\sin^2\theta_W$ is very close to 1/4 at the $M_Z$ scale is 
precisely the main reason for the suppression of the $Z'$-lepton 
couplings. Another important consequence of the former upper bound is 
that it leads to a constraint for the $SU(3)_L\otimes U(1)_X$ symmetry 
breaking scale $V$, and thus for the $Z'$ mass \cite{fra}. Indeed, taking 
into account the evolution of $\sin^2\theta$ with energy,  
it has been found \cite{ng} that the condition $\sin^2\theta(M_{Z'})\leq 
1/4$ imposes the upper limit $M_{Z'}\leq 3.1$ TeV. 
 
Let us concentrate now on the couplings between the fermions and 
the $Z'$, which from Eqs.~(\ref{lag}) and (\ref{rota}) come out to be 
\begin{equation} 
{\cal L}_{Z'} = - \frac{g}{\cos\theta_W} \left[\sum_{F=\psi,Q_i} 
\bar F_L\, g_L \,\gamma^\mu F_L + \sum_{q=u_i,d_i,J_i} \bar q_R\, 
g_R \,\gamma^\mu q_R \right] Z'_\mu 
\label{coup} 
\end{equation} 
where 
\begin{eqnarray} 
g_L & = & \frac{\sqrt{3}\,\sin^2\theta_W}{\sqrt{1-4\sin^2\theta_W}}\, X 
+ \sqrt{1-4\sin^2\theta_W}\; T_8 \nonumber \\ 
g_R & = & \frac{\sqrt{3}\,\sin^2\theta_W}{\sqrt{1-4\sin^2\theta_W}}\, X  
\end{eqnarray} 
Notice that when $\sin^2\theta_W$ approaches 1/4, the terms 
containing the $X$ generator enhance, while the $T_8$ part becomes 
suppressed. The key point is that the leptons are in an $X=0$ 
representation, therefore their couplings to the $Z'$ are solely given by 
the $T_8$ generator term. The quark couplings, on the contrary, will be 
dominated by the $X$ term, showing a relative enhancement of order 
$(1-4\sin^2\theta_W)^{-1}$. Another way of understanding this 
``leptophobia'' is just by looking at the rotation  
(\ref{rota}): in the $\sin^2\theta=1/4$ limit, we get $\cos\phi=0$, and 
the $B$ vector boson (corresponding to the $U(1)_X$) decouples, becoming 
the exact mass eigenstate identified with the $Z'$. As a consequence, its 
presence is not seen by the leptons, which are invariant under the 
$U(1)_X$ transformations. Of course, the mass decoupling is possible in 
this limit because the ratio $g'/g$ in (\ref{const}) diverges. 
 
\hfill 
 
{\em Effects on LEP observables and top quark production}. In order to 
analyse the $Z'$ effects 
within the model under consideration, let us first rewrite the couplings 
(\ref{coup}) as 
\begin{equation} 
{\cal L}_{Z'} = - \frac{g}{2\cos\theta_W} \sum_f \bar f\, (g'_{V_f} - 
g'_{A_f} \gamma_5)\,\gamma^\mu f\, Z'_\mu + \mbox{ flavor changing terms} 
\label{inter} 
\end{equation} 
where the sum is now carried out over all the fermions. We have followed 
here the notation in Refs.\ \cite{ver1,ver}, taking for the $Z'$-fermion 
couplings the same normalization as the usual one for the $Z$-fermion 
interactions within the SM, 
\begin{equation} 
{\cal L}_Z = - \frac{g}{2\cos\theta_W} \sum_f \bar f\, (g_{V_f} - 
g_{A_f} \gamma_5)\,\gamma^\mu f\, Z_\mu  
\end{equation} 
with $g_{V}=T_{3L}-2\,Q\sin^2\theta_W$, $g_A=T_{3L}$. Now, disregarding 
the exotic quarks $J_i$ (too massive to give any significant 
contribution), and using the definitions $h(x)\equiv\sqrt{1-4\,x}$, 
$x\equiv\sin^2 \theta_W$, the 3-3-1-model values for $g'_{V,A_f}$ are 
found to be 
\begin{equation} 
\begin{array}{lll} 
g'_{V_l} = \frac{3 h(x)}{2 \sqrt{3}}\; , & g'_{A_l} =  
- \frac{h(x)}{2 \sqrt{3}} & \mbox{for } l=e,\mu,\tau \\ 
g'_{V_\nu} = \frac{h(x)}{2 \sqrt{3}}\; , & g'_{A_\nu} =  
\frac{h(x)}{2 \sqrt{3}} & \mbox{for } \nu=\nu_e,\nu_\mu,\nu_\tau \\ 
g'_{V_{q^+}} = \frac{1}{\sqrt{3}\, h(x)}\, (-\frac{1}{2} + 3x + 
\delta_{q^+})\;, \hspace{.4cm} 
& g'_{A_{q^+}} = \frac{1}{\sqrt{3}\, h(x)}\, (- \frac{1}{2} - x + 
\delta_{q^+}) \hspace{.8cm} 
& \mbox{for } q^+=u,c,t \\ 
g'_{V_{q^-}} = \frac{1}{\sqrt{3}\, h(x)}\, (-\frac{1}{2} + 
\delta_{q^-})\; ,  
& g'_{A_{q^-}} = \frac{1}{\sqrt{3}\, h(x)}\, (- \frac{1}{2} + 2x + 
\delta_{q^-}) & \mbox{for } q^-=d,s,b 
\label{gs} 
\end{array} 
\end{equation} 
Notice the presence of the set of parameters $\delta_{q^\pm}$, which 
contain the non-universal part of the diagonal couplings in (\ref{inter}). 
This part, as well as the flavor changing terms, arises from the above 
mentioned ``discrimination'' between the quark families, thus the 
$\delta_{q^\pm}$ are related to the SM quark mixing matrix $V_{CKM}$ 
\cite{liu,tum}. One has 
\begin{equation} 
\delta_{q^\pm} = (1-x)\,\left[ V^{(\pm)\dagger} \mbox{ diag}(0,0,1) 
\, V^{(\pm)} \right]_{ii} 
\end{equation} 
where the index $i$ is equal to 1, 2 or 3 for $q^\pm$ belonging to the 
first, second or third quark generation respectively. The matrices 
$V^{(\pm)}$, $3\times 3$ and unitary, introduce new unknown parameters, 
since they are only constrained by the relation 
\begin{equation} 
V^{(+)\dagger}\, V^{(-)}=V_{CKM} 
\label{ckm} 
\end{equation} 
Their unitary character, however, is at least enough to ensure that the 
parameters $\delta_{q^\pm}$ lay in the range $0\leq \delta_{q^\pm} 
\leq 1-x$. 
 
Let us now turn to analyse the possibility of observation of the $Z'$ 
effects at LEP. For the allowed range of $M_{Z'}$, it can be seen that 
the effects from direct $Z'$ exchange are negligible at the $Z$ 
peak, thus the main contribution to the electroweak observables at LEP 
comes from the $Z-Z'$ mixing. In this way, the good agreement between 
the LEP measurements and the SM predictions for different asymmetries 
and decay rates serves as a constraint for the $Z'$ mass and the mixing 
angle $\beta$. With the 1992/93 available experimental data, such an 
analysis has been performed in Ref.\ \cite{ngliu}, leading to the 
bounds $-6\times 10^{-4} < \beta < 4.2\times 10^{-3}$ and $M_{Z_2} > 
490$ GeV, being $Z_2$ the exact mass eigenstate nearby the $Z'$. 
 
As stated above, further measurements found discrepancies with the SM 
values for the $Z\rightarrow b\bar b,$ $c\bar c$ decay rates, which 
could be attributed to this kind of $Z-Z'$ mixing. A model-independent 
analysis can be found in Ref.~\cite{ver1}, where the authors calculate the 
minimal allowed bands for the coupling constants $g'_{V,A_f}$ that are 
compatible with the LEP and SLC data. This analysis can be applied to 
the 3-3-1 model, where the lowest order $Z'$ contributions depend on 
the $Z'$ mass, the parameters $\delta_{b,c}$ and the mixing angle $\beta$. 
As expected, the $Z'$-quark couplings in this model show a significant 
enhancement with respect to the 
$Z$-quark ones in the SM, yielding for the ratios $\xi_{V,A_f}\equiv 
g'_{V,A_f}/g_{V,A_f}$ the values 
\begin{equation} 
\begin{array}{ll} 
\xi_{V_{q^+}} = 2.21 + 8.67 \,\Delta_{q^+} \hspace{3cm} & 
\xi_{A_{q^+}} = - 3.15 + 3.30 \,\Delta_{q^+} \\ 
\xi_{V_{q^-}} = 3.12 - 4.79 \,\Delta_{q^-} & 
\xi_{A_{q^-}} = 0.15 - 3.30 \,\Delta_{q^-} 
\label{xis} 
\end{array} 
\end{equation} 
which are obtained from (\ref{gs}) taking for $\sin^2\theta_W$ the 
measured effective value of 0.232. We have also normalized 
$\Delta_q\equiv\delta_q/(1-x)$, so that $0\leq\Delta_q\leq 1$. 
In terms of these ratios, the relative deviations of the widths 
$\Gamma(Z\rightarrow f\bar f)$ with respect to the SM predictions 
are given by 
\begin{equation} 
\frac{\delta\Gamma_{f\bar f}}{\Gamma_{f\bar f}} = \delta\rho \left( 
1\, + \,\frac{4x(1-x)}{1-2x}\; 
\frac{Q_f \,g_{V_f}}{g_{V_f}^2+g_{A_f}^2} \right) 
+ 2\,\beta\,\frac{g_{V_f}^2\,\xi_{V_f} + g_{A_f}^2\,\xi_{A_f}}{ 
g_{V_f}^2+g_{A_f}^2} 
\label{dgamma} 
\end{equation} 
where $\delta\rho$ stands for the mixing contribution to the 
$\rho$ parameter, defined as usual by $\rho\equiv M_W^2/(M_Z^2 
(1-x))$. For a mixing angle $\beta\ll 1$, it is easy to show that 
this contribution can be approximated by 
\begin{equation} 
\delta\rho \simeq \left( \frac{M_{Z'}}{M_Z} \right)^2 \beta^2 
\end{equation} 
Now, introducing in Eq.\ (\ref{dgamma}) the ratios in (\ref{xis}), 
we can estimate the relative deviations that correspond to the 
observables $R_b$, $R_c$ and $\Gamma_h$. These are found to be 
\begin{mathletters} 
\begin{eqnarray} 
\frac{\delta R_b}{R_b} & = & -0.06\, \delta\rho + 
[- 3.46 + 6.21 (\Delta_u+\Delta_c) ] \, \beta \\ 
\frac{\delta R_c}{R_c} & = & 0.12 \, \delta\rho + 
[-3.06 + 6.61 \Delta_c - 1.36 \Delta_u ] \, \beta \\ 
\frac{\delta \Gamma_h}{\Gamma_h} & = & 1.47\, \delta\rho + 
[ -1.88 + 1.36 (\Delta_u+\Delta_c) ] \, \beta 
\label{gh} 
\end{eqnarray} 
\label{rbrc} 
\end{mathletters} 
where we have taken into account the relations $\Delta_u+\Delta_c+ 
\Delta_t=1$ and $\Delta_d+\Delta_s+\Delta_b=1$, arising from the 
unitarity of the matrices $V^{(\pm)}$. We have also made use of 
the approximation $\Delta_b\simeq\Delta_t$, which is obtained from 
Eq.\ (\ref{ckm}) by neglecting the mixing angles between the third 
quark family and the other two in the $V_{CKM}$ matrix. 
 
It can be seen from (\ref{rbrc}) that a correction of a few per cent 
in the ratios $R_b$ and $R_c$ would require a mixing angle not far 
from the upper bound quoted above, $|\beta| \simeq 5\times 10^{-3}$. 
On the other hand, a relative deviation of $\Gamma_h$ below $0.2$\% 
(present experimental accuracy) would be possible either if the 
mixing angle satisfies $|\beta|\alt 1\times 10^{-3}$, or if we 
allow a cancellation between the different terms in (\ref{gh}). 
Since the mentioned observed deviations of $R_b$ and $R_c$ 
from the SM predictions are still under revision, we will not insist 
here on fitting the parameters of the model to the experimental results. 
However, it is worth to notice that the model-independent analysis in 
Ref.\ \cite{ver} suggest values of $|\xi_q|$ of about 3-4 in order 
to reproduce the possible excess of dijet events at CDF \cite{cdf} 
(the $Z'$ mass in this case is required to be around $800-900$ GeV). 
In the case of LEP, the presence of significant effects is also 
allowed, provided that the mixing angle $\beta$ has an absolute 
value of order $10^{-3}$ or higher. 
 
\hfill 
 
We proceed now to consider another potentially important effect of the 
presence of a $Z'$, namely the $Z'$ exchange contribution to the 
$t\bar t$ production cross section in high-energy $p\bar p$ collisions. 
This cross section has been experimentally measured by the CDF and 
D0 collaborations at Fermilab for a $p\bar p$ center-of-mass energy 
of $\sqrt{s}=1.8$ TeV, leading to the values $\sigma_{t\bar t}= 
7.5^{+1.9}_{-1.6}$ pb (CDF, $m_t=175$ GeV) \cite{scdf} and 
$\sigma_{t\bar t}=5.5\pm 1.8$ pb (D0, $m_t=173.3$ GeV) \cite{sdc}. 
The comparison with theoretical results shows that the value quoted 
by D0 is in good agreement with NLO QCD calculations, while that 
from CDF turns out to be somewhat higher than expected. Recent 
theoretical studies have been presented in Refs.\ \cite{mangano} 
and \cite{berger}, obtaining respectively $\sigma_{t\bar t}= 
4.75^{+0.73}_{-0.62}$  and $\sigma_{t\bar t}=5.52^{+0.07}_{-0.45}$ 
(in both cases $m_t=175$ GeV is considered). 
 
It has been argued \cite{gehr} that a $Z'$ having relatively large 
couplings to quarks (in particular, to the $u$ quark) could lead to 
significant corrections to $\sigma_{t\bar t}$. Indeed, these corrections 
have been explicitly evaluated in Ref.\ \cite{gehr} for a particular 
model containing a leptophobic $Z'$ \cite{alt}, showing that the effects 
can be important enough to 
be experimentally observable. We proceed here along the same lines, 
performing the relevant calculation for the $Z'$ boson within the 3-3-1 
model. Therefore, we consider the non-standard contribution from the 
$Z'$ exchange to the $q\bar q\rightarrow t\bar t$ 
annihilation subprocess, which at the leading order reads 
\begin{eqnarray} 
\hat\sigma(q\bar q\rightarrow Z'\rightarrow t\bar t) & = & 
\frac{(G_F M_Z^2)^2}{6 \pi} \;\frac{\hat s}{(\hat s-M_{Z'}^2)^2 
+(\hat s\,\Gamma_{Z'}/M_{Z'})^2} \nonumber \\ 
& & \times\, ({g_{V_q}'}^2+{g_{A_q}'}^2)\, \left[\frac{\beta_t}{2} 
\, (3-\beta_t^2)\, {g_{V_t}'}^2+\beta_t^3\, {g_{A_t}'}^2 \right] 
\end{eqnarray} 
where $\beta_t^2=1-4m_t^2/\hat s$, and $\Gamma_{Z'}$ is the total 
$Z'$ decay width. This 
parton-level expression has to be integrated over the corresponding 
parton distribution functions (PDF) 
in order to obtain the total $p\bar 
p$ cross section. We have carried out this calculation taking  
$m_t=175$ GeV, and using the MRS(A') PDF set  
\cite{mrs}, with $\alpha_s (M_Z^2)=0.113$. A $K$ factor of 
1.3 has also been included \cite{gehr} in order to 
account for the corresponding next-to-leading order corrections. 
 
Our results for different $Z'$ mass values are quoted in Table 
\ref{tabla}. We consider two possible $\Delta_q$ sets, (I) and 
(II), given by 
\begin{mathletters} 
\begin{eqnarray} 
(\mbox{I}) &\;\;\;& \Delta_t=\Delta_b=1\,,\;\; \Delta_q=0\;\mbox{ for 
other $q$} \\ 
(\mbox{II}) & & \Delta_c=1\,,\;\; \Delta_s=\cos^2\theta_c\,,\;\; 
\Delta_d=\sin^2\theta_c\,,\;\; \Delta_b=\Delta_u=\Delta_t=0 
\end{eqnarray} 
\end{mathletters} 
where $\theta_c$ is the Cabbibo angle. For the quoted $M_{Z'}$ values, 
these sets provide with good approximation the highest (set I) and lowest 
(set II) contributions to $\sigma_{t\bar t}$ within the allowed 
parameter space. We find that a $Z'$ having a mass 
lower than 700 GeV appears to be excluded, while a $Z'$ heavier than 
900 GeV would need an experimental accuracy of 1 pb or less in 
$\sigma_{t\bar t}$ in order to produce a visible effect. In the remaining 
range, i.e. $M_{Z'}=700$ to 900 GeV, the uncertainty introduced by the 
$\Delta_q$ parameters, together with the discrepancy between the 
experimental values obtained by CDF and D0, do not allow a conclusive 
analysis. However, the above results from both experiments only consider 
channels with at least one lepton, and the experimental situation can 
be improved by the inclusion of all the hadronic 
channels. Then a further reduction in the uncertainty (up to a 10\%) is 
expected for the second run, by the year 2001 \cite{scdf}. 
 
\hfill 
 
{\em Effects at LEP2.} We concentrate now on the possible $Z'$ effects 
on the experimental quantities that will be measured at LEP2. The 
observables of interest in this case are those  
related to the leptonic cross sections $e^+ e^-\rightarrow l^+ l^-$,  
together with three hadronic quantities, namely the $e^+ e^- 
\rightarrow b\bar b$ cross section $\sigma_b$, the total hadronic cross 
section $\sigma_h$ and the forward-backward asymmetry in the $b\bar b$ 
production $A_{FB,b}$. We see that the presence of a $Z'$ would give 
rise to deviations from the SM predictions for these observables, which 
turn out to be negligible for the leptonic channels, but not necessarily 
for the mentioned hadronic quantities. 
 
By considering the $Z'$ contributions to the {\it leptonic} observables, 
it is possible to determine a region in the $(g'_{V_l},g'_{A_l})$ plane 
where the predictions of the SM cannot be distinguished at LEP2 from those 
of the models including the extra $Z$. A model-independent analysis 
\cite{lei} shows that (with 95\% CL) this region is bounded by 
\begin{equation} 
|g'_{V_l}| \alt 0.12\,\sqrt{\frac{M_{Z'}^2 - s}{s}} 
\hspace{3cm} 
|g'_{A_l}| \alt 0.096\,\sqrt{\frac{M_{Z'}^2 - s}{s}} 
\label{lept} 
\end{equation} 
where $s$ stands for the squared $e^+ e^-$ center-of-mass energy. In the 
case of the 3-3-1 model, the values for $g'_{V,A_l}$ contain no free  
parameters, thus the relations (\ref{lept}) can be 
unambiguously checked for given values of $s$ and the $Z'$ mass. With 
$\sin^2\theta_W=0.232$, we get from (\ref{gs}) 
\begin{equation} 
g'_{V_l}\simeq 0.23 \hspace{4cm} g'_{A_l}\simeq -0.08 
\label{gl} 
\end{equation} 
Therefore, if the configuration $s_0 = (175 \mbox{ GeV})^2$ (the most 
convenient one \cite{lei}) is chosen, it is seen that the values in 
(\ref{gl}) lay within the above non-observability region, even for 
a  ``light'' $Z'$ of mass $\alt 700$ GeV. 
 
The situation is more promising for the above mentioned {\it hadronic} 
observables, although in this case the $Z'$ couplings depend on  
the unknown parameters $\Delta_q$. In particular, we find that the 
total hadronic cross section could be increased by about 2\% 
even for a $Z'$ mass of $\sim 900$ GeV. In order to estimate the  
shifts for the hadronic observables with respect to the SM 
predictions, we use here the so-called ``$Z$-peak subtracted'' 
approach \cite{ren}, which improves the Born approximation by taking 
measured $Z$-peak quantities as input parameters (for the case of a 
$Z'$ having universal couplings to quarks, a similar analysis has 
already  been performed in Refs.\ \cite{ver1,ver}). In this way, 
the relative corrections for $s=s_0$ in the 3-3-1 model are found to be 
\begin{eqnarray} 
\frac{\delta \sigma_b}{\sigma_b} & = & \left[ 5 \times 10^{-2} \right] 
\left\{ \frac{(1 \mbox{ TeV})^2}{M_{Z'}^2 - s_0} \; \left[ 0.18 - 0.13 
\, \Delta_b \right] \right\} 
\nonumber \\  
\frac{\delta \sigma_h}{\sigma_h} & = & \left[ 1.4 \times 10^{-2} \right] 
\left\{\frac{(1 \mbox{ TeV})^2}{M_{Z'}^2 - s_0} \; \left[ 0.65 + 0.35 \, 
(\Delta_u + \Delta_c) \right] \right\} 
\nonumber \\  
\frac{\delta A_{FB,b}}{A_{FB,b}} & = & \left[ 10 \times 10^{-2} \right] 
\left\{\frac{(1 \mbox{ TeV})^2}{M_{Z'}^2 - s_0} \; \left[ 0.19 - 0.59 \, 
\Delta_b \right] \right\} 
\label{deltas} 
\end{eqnarray} 
where once again the relation $\Delta_d+\Delta_s+\Delta_b=1$ has been 
used. For each observable, the corresponding accuracy (two standard 
deviations) expected at LEP2 has been explicitly factorized out, 
so that the expressions in curly brackets have to be at least of order 
one to yield visible effects. It is seen from (\ref{deltas}) that 
the uncertainty introduced by the parameters $\Delta_q$ 
is minimal for $\sigma_h$, where a positive enhancement can be obtained. 
In fact, in this case the shift from the SM value can be produced 
even at three standard deviations for a $Z'$ mass below 900 GeV. For the 
two other observables, the situation is less favourable, although the 
use of the combined observables would allow to set improved negative 
constraints (recall that $\Delta_u+\Delta_c\simeq 1-\Delta_b$).  
 
Taking now 
into account the above results from the analysis of the effects 
on $t\bar t$ production, we conclude that a $Z'$ having a mass in the 
range $700-900$ GeV is likely to produce significant contributions to 
both the top production cross section $\sigma_{t\bar t}$ measured at  
Fermilab and the total hadronic cross section $\sigma_h$ to be  
measured at LEP2. The shift from the SM predictions appears to be 
particularly important for the latter, where the effect of the  
unknown parameters turns out to be minimized and a visible signal can 
be obtained even for $M_{Z'}$ up to 1 TeV. Since the hadronic cross 
section is expected to be measured with relatively good accuracy at LEP2, 
we believe that the 3-3-1 model should be considered as a potentially 
interesting one in the forthcoming runs, when the expected luminosity 
will be achieved.  
 
\acknowledgements 
 
The author is indebted to D.\ Comelli, L.\ Epele, C.\ Garc\'{\i}a 
Canal and C.\ Verzegnassi for useful discussions. This work has 
been partially supported by a research fellowship from the 
Universitat de Val\`encia (Spain), and a grant from the Commission 
of the European Communities, under the TMR programme (Contract 
N$^\circ$ ERBFMBICT961548). Financement has also 
been received from CICYT, Spain, under Grant AEN-96/1718.

\begin{table} 
\caption{Contributions to $\sigma_{t\bar t}$ from the extra $Z$ boson, 
for $\Delta_q$ sets (I) and (II) and different values of the $Z'$ mass.} 
 
\begin{tabular}{ccccc} 
\hspace{0cm} & $M_{Z'}$ (GeV) & $\sigma_{t\bar t}^{(I)}$ (pb) & 
$\sigma_{t\bar t}^{(II)}$ (pb) & \hspace{0cm} \\ 
\hline 
& 700 & 4.3 & 1.7 & \\ 
& 800 & 1.9 & 0.77 & \\ 
& 900 & 0.82 & 0.34 & \\ 
& 1000 & 0.39 & 0.15 & \\ 
& 1200 & 0.11 & 0.04 & 
\end{tabular} 
\label{tabla} 
\end{table} 
 

\begin{references} 
\bibitem{ver1} P.\ Chiapetta, J.\ Layssac, F.\ M.\ Renard and C.\ 
Verzegnassi, Phys.\ Rev.\ D {\bf 54} (1996) 789. 
 
\bibitem{alt} G.\ Altarelli, N.\ Di Bartolomeo, F.\ Feruglio, R.\ Gatto 
and M.\ L.\ Mangano, Phys.\ Lett.\ B {\bf 375} (1996) 292. 
 
\bibitem{lep} The LEP collaborations ALEPH, DELPHI, L3, OPAL and the LEP 
Electroweak Working Group, preprint CERN-PPE/95-172. 
 
\bibitem{cdf} CDF Collaboration, F.\ Abe {\em et al.}, Phys.\ Rev.\ Lett.\ 
{\bf 77} (1996) 438. 
 
\bibitem{ver} G.\ Montagna, F.\ Piccinini, J.\ Layssac, F.\ M.\ Renard and 
C.\ Verzegnassi, preprint PM/96-25, FNT/T-96/18, {\tt hep-ph/9609347}. 
 
\bibitem{pol}A.\ Blondel, {\em Status of the electroweak interactions, 
experimental aspects}, talk given at the ICHEP 96, Warsaw, Poland, to 
appear in the proceedings.  
 
P.\ M\'el\`ese, for the CDF Collaboration, FERMILAB-CONF-97/167-E. 
Presented at the 11th Les Rencontres de Physique de la Vallee d'Aoste, 
La Thuile, Italy, March 1997. 
 
\bibitem{otros} K.\ S.\ Babu, C.\ Kolda and J.\ March-Russell, Phys.\ 
Rev.\ D {\bf 54} (1996) 4635. 
 
K.\ Agashe, M.\ Graesser, I.\ Hinchliffe and M.\ Suzuki, Phys.\ Lett.\ B 
{\bf 385} (1996) 218; H.\ Georgi and S.\ Glashow, Phys.\ Lett.\ B {\bf 387} 
(1996) 341. 
 
P.\ H.\ Frampton, M.\ B.\ Wise and B.\ D.\ Wright, Phys.\ Rev.\ D {\bf 54} 
(1996) 5820. 
 
\bibitem{vic} F.\ Pisano and V.\ Pleitez, Phys.\ Rev.\ D {\bf 46} (1992) 410. 
 
\bibitem{fra} P.\ H.\ Frampton, Phys.\ Rev.\ Lett.\ {\bf 69} (1992) 2889. 
 
\bibitem{ng} D.\ Ng, Phys.\ Rev.\ D {\bf 49} (1994) 4805. 
 
\bibitem{cp} L.\ Epele, H.\ Fanchiotti, C.\ Garc\'{\i}a Canal and D.\ 
G\'omez Dumm, Phys.\ Lett.\ B {\bf 343} (1995) 291. 
 
\bibitem{foot} R.\ Foot, O.\ F.\ Hernandez, F.\ Pisano and V.\ Pleitez, 
Phys.\ Rev.\ D {\bf47} (1993) 4158. 
 
\bibitem{liu} J.\ T.\ Liu, Phys.\ Rev.\ D {\bf 50} (1994) 542. 
 
\bibitem{tum} D.\ G\'omez Dumm, F.\ Pisano and V. Pleitez, Mod.\ Phys.\ 
Lett.\ A {\bf 9} (1994) 1609.\  
 
\bibitem{ngliu} J.\ T.\ Liu and D.\ Ng, Zeit.\ Phys.\ C {\bf 62} (1994) 
693.  
 
\bibitem{scdf} A.\ Castro, for the CDF and D0 Collaborations,  
FERMILAB-CONF-97/137-E. Presented at the 32nd Rencontres de Moriond, 
Les Arcs, France, March 1997. 
 
\bibitem{sdc} D0 Collaboration, S.\ Abachi {\em et al.}, preprint 
FERMILAB-Pub-97/109-E, {\tt hep-ph/ 9704015}. 
 
\bibitem{mangano} S.\ Catani, M.\ L.\ Mangano, P.\ Nason and L.\ 
Trentadue, Phys.\ Lett.\ B {\bf 378} (1996) 329. 
 
\bibitem{berger} E.\ L.\ Berger and H.\ Contopanagos, Phys.\ Rev.\ D 
{\bf 54} (1996) 3085. 
 
\bibitem{gehr} T.\ Gehrmann and W.\ J.\ Stirling, Phys.\ Lett.\ B 
{\bf 381} (1996) 221. 
 
\bibitem{mrs} A.\ D.\ Martin, W.\ J.\ Stirling and R.\ G.\ Roberts, 
Phys.\ Lett.\ B {\bf 354} (1995) 155. 
 
\bibitem{lei} P.\ Chiappetta {\it et al.}, in {\it Physics at LEP2}, CERN 
Report 96-01, Eds.\ G.\ Altarelli, T.\ Sj\"ostrand and F.\ Zwirner, Vol.\ 
I, p.\ 577. 
\bibitem{ren} F.\ M.\ Renard and C.\ Verzegnassi, Phys.\ Rev.\ D {\bf 53}  
(1996) 1290. 
 
\end{references}
\end{document}